\documentclass[12pt]{article}
\usepackage[utf8]{inputenc}
\usepackage{empheq}
\usepackage{bm}
\usepackage[normalem]{ulem}
\usepackage{latexsym}
\usepackage{color}
\usepackage{amsmath,amsfonts,amssymb}
\usepackage{graphicx,epsfig}
\usepackage{psfrag}
\usepackage{amsthm}
\usepackage{cite,url}
 \usepackage{makecell}

\pdfoutput=1

\usepackage{amsmath}
\usepackage{amssymb}
\usepackage{amsfonts}
\usepackage{mathrsfs}

\usepackage{mathrsfs}
\usepackage{fullpage}
\usepackage{setspace}
\usepackage{bm}
\usepackage{bbm}

\usepackage[normalem]{ulem}
\usepackage{enumerate}
\usepackage{yfonts}

\usepackage{psfrag}

\usepackage{graphicx}
\usepackage{cancel}
\usepackage{slashed}

\usepackage[font=small,labelfont=bf]{caption}
\usepackage{subcaption}

\newcommand{\be}{\begin{equation}}
\newcommand{\bea}{\begin{eqnarray}}
\newcommand{\ee}{\end{equation}}
\newcommand{\eea}{\end{eqnarray}}

\begin{document}
\numberwithin{equation}{section}
{
\begin{titlepage}
\begin{center}

\hfill \\
\hfill \\
\vskip 0.2in

{\Large Study of eccentric binaries in Horndeski gravity}\\

\vskip .7in

{\large
Abhishek Chowdhuri${}$\footnote{\url{chowdhuri_abhishek@iitgn.ac.in
}}, Arpan Bhattacharyya${}$\footnote{\url{abhattacharyya@iitgn.ac.in}}
}

\vskip 0.3in

{\it Indian Institute of Technology, Gandhinagar, Gujarat-382355, India}\vskip .5mm

\vskip.5mm

\end{center}

\vskip 0.35in

\begin{center} {\bf ABSTRACT } \end{center}
We study the orbital evolution of eccentric binary systems in Horndeski gravity. This particular theory provides a testbed to give insightful comparisons with data. We compute the rate of energy loss and the rate of change of angular momentum for the binaries by calculating the multipole moments of the radiation fields. We have used appropriate parameters for the eccentric binaries to compute the decay rates of its orbital eccentricity and semi-major axis. We then compare this decay rate with that of GR.
\vfill


\end{titlepage}
}

\newpage
\tableofcontents
\section{Introduction}
General Relativity is widely considered to be the most successful theory of classical gravity. It is so because it qualifies tests ranging from solar system, pulsar tests to sub-millimeter scales\cite{Hoyle:2000cv,2001CQGra..18.2397A,1993tegp.book.....W,Will:2014kxa,Stairs:2003eg,Wex:2014nva,Manchester:2015mda,Kramer:2016kwa}. However, it is incomplete. It is widely expected that there will be a UV completion of the theory by constructing a suitable quantum theory\cite{Keifer}. Nevertheless, GR, along with Standard Model being the two cornerstones of physics, the GR theory itself cannot explain all phenomena there is, the biggest instance being the occurrence of dark energy and dark matter\cite{Cline}. Supernovae Type IA observations \cite{SupernovaSearchTeam:1998fmf,SupernovaSearchTeam:2004lze,SupernovaCosmologyProject:1998vns,WMAP:2006bqn} show that our universe is expanding. To explain this in the context of GR, one introduces an extra piece, ``Dark energy", which is not understood completely. Dark matter is yet another mysterious thing ever-present in galaxies and taking part in gravitational interactions\cite{Oort,Zwicky,Bertone:2004pz}. All these questions naturally seek out an attempt to investigate alternate gravity theories\cite{Yunes:2013dva,Clifton:2011jh}. 
\par
 One way we can modify GR is to add terms to it which are higher curvature terms and also add in additional degrees of freedom\cite{Alexeev:1996vs,Alexeyev:2006zz,Sullivan:2019vyi,Cano:2019ore,Lehebel:2018zga,Volkov:2016ehx,Arun,Kunz:2006ca}. The simplest way is to add scalars\cite{Maeda,Faraoni,Damour:1992we,Horbatsch:2015bua,Schon:2021pcv,Rainer:1996gw}. Such theories with extra scalars do give solutions for some GR problems. For instance, the scalar field is considered to be one of the main components of dark energy and can also explain the reason for the accelerating expansion of the universe\cite{DeFelice:2011bh,Gsponer:2021obj}. In this context, Horndeski gravity presents itself to be one of the most general of such theories evading instabilities of the Ostrogradsky type\cite{Horndeski,Kobayashi:2019hrl}. The theory has been extensively tested in cluster lensing experiments\cite{Narikawa:2013pjr,Zhang:2021ygh}, CMB data\cite{Salvatelli:2016mgy,Renk:2017rzu,Heisenberg:2022lob,Heisenberg:2022gqk,Lee:2022cyh}, etc and widely studied in the context of cosmology\cite{Germani:2017pwt,Kennedy:2017sof,Santos:2019ljs,Tretyakova:2017rlk,Tretyakova:2017fbu}. However, on the question of its verification using LIGO-Virgo data\cite{Abbott:2016blz,TheLIGOScientific:2016wfe,Abbott:2016nmj,Abbott:2017vtc,LIGOScientific:2018mvr,LIGOScientific:2020ibl,LIGOScientific:2021djp}, some recent works do restrict some parameters of the theory using the event GW170817\cite{LIGOScientific:2017vwq,LIGOScientific:2017zic}.
\par
In the Horndeski theory, there is a coupling between the scalar and the matter fields. 
If we have scalar fields describing dark energy, then there exists a mechanism called Vainshtein mechanism\cite{Vainshtein:1972sx}, responsible for suppressing the scalar interactions. Current experimental constraints also require this interaction to be screened in high-density medium\cite{Adelberger:2009zz,Williams:2012nc}. One can define  a radius, known as  `Vainshtein radius' beyond which one can neglect all the non-linearities arising due to the presence of higher derivative interactions\cite{Khoury:2003rn,Khoury:2003aq,Hinterbichler:2010es,Hinterbichler:2011ca,Damour:1994zq,Damour:1994ya}. 
In this paper, we consider a Horndeski theory, where this mechanism is neglected for binary pulsars in the strong field regime, and the scalar field equations reduce to a massive Klein-Gordon equation\footnote{A massless version of this theory has been considered \cite{Hou:2017cjy} where authors have computed time delay, Nordtvedt effects, etc.}.
\par
 The first detection of gravitational waves was an indirect one based on orbital decays of binary pulsars\cite{Taylor:1994zz}, and it was only in 2015 we found a direct signal from the coalescence of two stellar-mass black holes\cite{Abbott:2016blz}. GWs provides us with an opportunity to study dynamics and hence perform quantitative tests. The study of radiations clarifies predictions from GR and the alternative theories of gravity. Therefore, studying orbital radiations is an important task, which is what we have done here. Gravitational fields for pulsars provide a strong test bed than the usual solar system tests. Also, there's a high stability for pulse arrivals so that we can minutely study the orbital dynamics of such systems\cite{Hobbs:2019ktp,Ivanov:2016ifg}. These facts suffice to put pulsars as a particular interesting thing to study gravitational radiations. We can do so in different models of gravity, and some of them have been done in \cite{Pshirkov:2008nr}. In \cite{Dyadina:2018ryl}, authors have considered circular orbits for binaries and constrained the effective mass of the scalar field using some binary pulsar data. However, we know that there are eccentric binaries pulsars\cite{Freire:2012mg} \footnote{Also, we expect to detect gravitational waves from eccentric binary systems in future detectors like LISA \cite{Fang:2019dnh}. Hence we expect to get better constraints on the parameter space of the theory.}. Motivated by that, in this paper, we study decay rates corresponding to energy and angular momentum for binary pulsar systems moving in eccentric orbits, with the hope that using these results, one can possibly improve the bound on the parameters of the theory. We compute the angular momentum flux from the method outlined in \cite{Isaacson:1968hbi,Isaacson:1968zza} and use a Keplerian parametrization for the pulsar systems. We conclude by finding the expressions for the rate of change of eccentricity of the orbits of the binaries. 
 
\par
 We organize the paper as follows: In Sec.~(\ref{sec1}), we give a brief introduction to the Horndeski theory. In Sec.~(\ref{sec4}), we look into the field equations of motion and investigate their weak limits. Sec.~(\ref{sec5}) deals with finding the equation of motion for the binaries in this particular gravity theory,  followed by Sec.~(\ref{5}), which outlines the way we use to calculate the stress-tensor. We find solutions to the field equations in Sec.~(\ref{sec6}) which, along with the stress tensor, helps in calculating the energy and the angular momentum flux in Sec.~(\ref{sec7}). Finally, we analyze the orbital dynamics for such systems in Sec.~(\ref{sec8}). We conclude with Sec.~(\ref{sec9}) by providing a summary of our results and future directions. Some mathematical details are given in the Appendix~(\ref{A}).
\par
\textbf{Conventions:} Choice of signature for the metric is (-,+,+,+). We use Greek indices $(\mu,\nu,..)$ running over 0,1,2,3, and the calculations are done in the CGS system.


 \section {Horndeski Gravity}\label{sec1}
 The action functional for the Horndeski theory is given by
 \cite{Kobayashi:2019hrl,Kobayashi:2011nu}
 \begin{equation} \label{Horn}
 S=\frac{c^{4}}{16\pi} \sum_{i=2}^{5}\int d^{4}x \sqrt{-g}L_{i}+\int d^{4}x\, L_{m}(A^{2}(\phi)g^{\mu \nu},\psi_{m}^{(j)})
 \end{equation}
 where c is the speed of light, g is the determinant of the metric, and $L_{m}$ is the matter part of the Lagrangian, the matter fields $\psi_{m}^{j}$, being labelled by j. $L_{i}$'s are the  Lagrangian densities for the gravity part containing the metric and the scalar field $\phi.$
\begin{align}
\begin{split}
& L_{2}=G_{2}(\phi,X), L_{3}=-G_{3}(\phi,X)\Box \phi,L_{4}=G_{4}(\phi,X)R+G_{4X}[(\Box \phi)^{2}-(\nabla_{\mu}\nabla_{\nu}\phi)^{2}],\\& L_{5}=G_{5}(\phi,X)G_{\mu \nu}\nabla^{\mu}\nabla^{ \nu}\phi-\frac{G_{5X}}{6}[(\Box \phi)^{3}+2(\nabla_{\mu}\nabla_{\nu}\phi)^{3}-3(\nabla_{\mu}\nabla_{\nu}\phi)^{2}\Box \phi].
\end{split}
\end{align}
$G_{\mu \nu}$ is the usual Einstein tensor, $R$ is the Ricci scalar. The `$X$' is defined as $$X=-\frac{1}{2}\nabla_{\mu}\phi \nabla^{\mu}\phi.$$ $G_{i}(\phi,X)$ are functions of the scalar field $\phi$ and $X.$ Furthermore, $G_{iX}=\frac{\partial G_{i}}{\partial X}$.
\par 
Since we are interested in studying gravitational radiations from binary pulsar systems, we will concentrate on the flux carried away by these radiations at large distances. At large distances influences on the metric and scalar field $\phi$ are not significant from these pulsar sources. Hence we can perturb the field equations in a background, which is Minkowskian ($\eta_{\mu\nu}$) and expand the fields in the limit of $\frac{v}{c} \ll 1$ \cite{Poisson,Will:2014kxa}. The metric and the scalar field expansions are given by,
\begin{align}
    \begin{split}
& \phi=\phi_{0}+\tilde{\phi}\,,\\ & g_{\mu \nu}=\eta_{\mu \nu}+h_{\mu \nu}.
    \end{split}
\end{align}
In the subsequent sections we will solve these perturbations in terms of power series of  $\mathcal{O}(v/c)^{2}$, using the standard Post-Newtonian techniques.\par
The functions $G_{i}(\phi,X)$ are arbitrary and meanwhile can be expanded in Taylor's series around $\phi_{0}$ which is the value of $\phi$ in flat spacetime background.
\begin{align}
    \begin{split} 
        & G(\phi, X)=\sum_{m,n=0}^{\infty}G_{(m,n)}\tilde{\phi}^{m}X^{n}\,, \\ & G_{(m,n)}=\frac{1}{m!n!}\frac{\partial^{m+n}}{\partial^{m}\phi \partial^{n}X}G(\phi,X)|_{\phi=\phi_{0},X=0}\,,\\ & G(\phi,X)=G(\phi)|_{n=0}. \label{GG}
    \end{split}
\end{align}
To obtain the GR limit of this theory we can choose $G_{4}=\frac{1}{G_{N}}$ and set all other $G_{i}=0$. Taking into account existing constraints posed on the theory from various observational data we reduce our original Lagrangian (\ref{Horn}) to the following \cite{Dyadina:2018ryl},
\begin{equation}
    L_{2}=G_{2}(\phi,X),L_{3}=-G_{3}(\phi)\Box \phi, L_{4}=G_{4}(\phi)R, L_{5}=0.
\end{equation}
However, there are still some undetermined parameters left which one hope to constrain. Using one results of this paper we hope to get back to this issue in a subsequent work.
\par
Now, let's concentrate on the matter part of the theory. Since we are studying gravitational pointlike masses, the matter action for such a system is given by \cite{eardley},
\begin{equation} \label{mact}
    S_{m}=-c^{2}\sum_{a}\int m_{a}(\phi)d\tau_{a}.
\end{equation}
 Here, $m_{a}(\phi)$ is the inertial mass of the particles and $\tau_{a}$ is the proper time along the world line of the particle, $a$ being a label denoting various point particles. As we will consider a binary system, it will take only two values $a=1,2.$ The above action was first considered in the context of scalar-tensor theories to account for the internal structure of the bodies\cite{eardley}. This is so because, for alternate gravity theories like scalar-tensor theories, laws of self-gravitating bodies depend on their internal structure, and there is a violation of the Gravitational Weak Equivalence Principle (GWEP)\cite{DiCasola:2013iia,Damour}. The violation can be easily seen as the scalar field and hence the mass which depends on the field becomes position-dependent. At this point, we like to emphasize that, instead of a generic matter Lagrangian $L_m$ as mentioned (\ref{Horn}), we will be working with the simplified matter action as mentioned in (\ref{mact}). This is a valid approximation as long as the compact objects are far from each other. Their motion can be effectively described by assuming that they are point particles. Then the effect of the scalar field on the internal structure of these objects can be expressed by making their mass a function of the scalar field as shown in (\ref{mact}) following the arguments given in \cite{eardley,Dyadina:2018ryl, Hou:2017cjy}. \par
 
 Varying (\ref{mact}) with respect to the metric we obtain 
the energy-momentum tensor of the matter \cite{Dyadina:2018ryl}
\begin{align}
    \begin{split} \label{stressexpr}
& T^{\mu \nu}=\frac{c}{\sqrt{-g}}\sum_{a}m_{a}(\phi)\frac{u^{\mu}u^{\nu}}{u^{0}}\delta^{3}({\vec{r}-\vec{r}_{a}(t)})\,,
    \end{split}
    \end{align}
    and its trace is given by
    \begin{align}
        \begin{split}
             &  T=-\frac{c^{3}}{\sqrt{-g}}\sum_{a}\frac{m_{a}(\phi)}{u^{0}}\delta^{3}(\vec{r}-\vec{r}_{a}(t))
        \end{split}
    \end{align}
    where, $u^{\mu}=\frac{dx_{a}^{\mu}}{d\tau_{a}}$ is the four velocity of the a-th particle, $d\tau=\frac{\sqrt{-ds^{2}}}{c}$, $ds^{2}=g_{\mu \nu}dx^{\mu}dx^{\nu}$ is an interval, $u_{\mu}u^{\mu}=-c^{2}$ and $\delta^{3}(\vec{r}-\vec{r}_{a}(t))$ is the three dimensional Dirac delta function. 
  The inertial mass $m$ which is dependent on $\phi$, can also be expanded around $\phi_{0}$ and takes the form \cite{Dyadina:2018ryl},
\begin{equation}
    m_{a}(\phi)=m_{a}(\phi_{0})\left[1+s_{a}\frac{\tilde{\phi}}{\phi_{0}}-\frac{1}{2}\left(\frac{\tilde{\phi}}{\phi_{0}}\right)^{2}(s_{a}-s_{a}^{2}-s_{a}^{\prime})+O(\tilde{\phi}^{3})\right] \label{mphi}
\end{equation}
where $s_{a}=\frac{\partial (\ln m_{a})}{\partial(\ln \phi)}|_{\phi_{0}}$ and $ s_{a}^{\prime}=\frac{\partial^{2}(\ln m_{a})}{\partial (\ln \phi)^{2}}|_{\phi_{0}}.$\\

Having obtained these expressions the field equations can be expressed in the weak field limit which we will do in the next section.

\section{Weak field limits} \label{sec4}
As mentioned earlier, the scope of this paper is to explore radiations from a pulsar system at a large distance from the source such that there is no influence of the source on the metric and scalar field. Hence, the following perturbations can be performed
\begin{align}\label{3.1}
    \begin{split}
& \phi=\phi_{0}+\tilde{\phi}\,,\\ & g_{\mu \nu}=\eta_{\mu \nu}+h_{\mu \nu}.
    \end{split}
\end{align}
This leads to a Taylor series expansion of the arbitrary function $G_{i}(\phi,X)$ defined in (\ref{GG}). 
\par
The next thing to do is to obtain field equations in the above limit, known the weak field limit. Interested readers may refer to\cite{Kobayashi:2011nu,Gao:2011qe} for the most general form of the field equations. Furthermore, following \cite{Ashtekar:2015ooa} we have neglected  $G_{2(0,0)}$ and $G_{2(1,0)}$ as these terms have negligible effects on gravitational wave. The linearized field equations upto $\mathcal{O}(\frac{v}{c})^{4}$ takes the following form \cite{Hohmann:2015kra},
\begin{align}
    \begin{split}
        & -(G_{2(0,1)}-2G_{3(1,0)})\Box \tilde{\phi}-G_{2(2,0)}\tilde{\phi}+G_{4(1,0)}(\Box h-\partial_{\mu}\partial_{\nu}h^{\mu \nu})=\frac{16 \pi}{c^{4}}\frac{\partial T}{\partial \phi}^{(1)}\,, \\ & G_{4(0,0)}(\partial_{\alpha}\partial_{\nu}h^{\alpha}_{\mu}-\frac{1}{2}\Box h_{\mu \nu}-\frac{1}{2}\partial_{\mu}\partial_{\nu}h-\frac{1}{2}\eta_{\mu \nu}\partial_{\alpha}\partial_{\beta}h^{\alpha \beta}+\frac{1}{2}\eta_{\mu \nu}\Box h)\\ & +G_{4(1,0)}\eta_{\mu \nu}\Box \tilde{\phi}-G_{4(1,0)}\partial_{\mu}\partial_{\nu}\tilde{\phi}=\frac{8\pi}{c^{4}}T_{\mu \nu}^{(1)} 
    \end{split}
\end{align}
where $T=g^{\mu \nu}T_{\mu \nu}$ is the trace, $\Box=\eta^{\mu \nu}\partial_{\mu}\partial_{\nu} $ and the the superscript $(1)$ refers to the leading order term. 
\par
By introducing,
\begin{align}
    \begin{split} \label{3.3}
        & \tilde{h}_{\mu \nu}=h_{\mu \nu}-\frac{1}{2}\eta_{\mu \nu}h-\frac{G_{4(1,0)}}{G_{4(0,0)}}\eta_{\mu \nu}\tilde{\phi}\,, \\ & \tilde{h}=-h-4\frac{G_{4(1,0)}}{G_{4(0,0)}}\tilde{\phi}\,,
    \end{split}
\end{align}
the field equations can be decoupled and further using the gauge $\partial_{\alpha}h^{\alpha \beta}=0$, the field equations take the following form,  
\begin{equation} \label{3.4}
    \Box \tilde{h}_{\mu \nu}=-\frac{16 \pi}{c^{4}G_{4(0,0)}}T_{\mu \nu}^{(1)}\,,
\end{equation}
\begin{equation} \label{3.5}
    \Box \tilde{\phi}-m_{s}^{2}\tilde{\phi}=\frac{16 \pi}{c^{4}}\xi\left[T^{(1)}-\frac{2G_{4(0,0)}}{G_{4(1,0)}}\frac{\partial T}{\partial \phi}^{(1)}\right]\,,
\end{equation}
where 
\begin{equation}
   m_{s}^{2}=\frac{G_{2(2,0)}}{2G_{3(1,0)}-G_{2(0,1)}-3\frac{G^{2}_{4(1,0)}}{G_{4(0,0)}}} \,,
\end{equation}
\begin{equation}
    \xi=-\frac{G_{4(1,0)}}{2G_{4(0,0)}\left(2G_{3(1,0)}-G_{2(0,1)}-3\frac{G^{2}_{4(1,0)}}{G_{4(0,0)}}\right)}.
\end{equation}
In this paper, we have chosen CGS system of units which makes the dimension of $m_{s}$ inverse of the length and consequently it plays a role of inverse characteristic wavelength of the scalar field. The scalar field, whether it is massless or massive depends on the value of $G_{2(2,0)}$ and the perturbation of the scalar field is sourced by a combination of energy-momentum stress tensor of the matter and the derivative of the trace of the tensor with respect to $\phi$. This is however not the case for the equation of $\tilde{h}_{\mu\nu}$, where the energy-momentum tensor of the matter field is sourcing the perturbation.
\par
Now using the expansions mentioned in (\ref{3.1}) and the fact that $u_{\mu}u^{\mu}=-c^{2},$ the energy-momentum tensor takes the following form after expanding the expression mentioned in (\ref{stressexpr}) upto the linear order,
\begin{equation}
    T^{\alpha \beta}=\sum_{a}m_{a}u^{\alpha}u^{\beta}\left(1-\frac{1}{2}h^{k}_{k}-\frac{v_{a}^{2}}{2c^{2}}+s_{a}\frac{\tilde{\phi}}{\phi_{0}}\right)\delta^{3}(\vec{r}-\vec{r}_{a}(t)).
\end{equation}
Note that we have to use also the expression of $m(\phi)$ as defined in (\ref{mphi}), and we have neglected the quadratic term in (\ref{mphi}) as we are considering fluctuations as mentioned in (\ref{3.1}).
The trace of $T^{\alpha\beta}$ is, 
\begin{equation}
    T=-c^{2}\sum_{a}m_{a}\left(1-\frac{1}{2}h^{k}_{k}-\frac{v_{a}^{2}}{2c^{2}}+s_{a}\frac{\tilde{\phi}}{\phi_{0}}\right)\delta^{3}(\vec{r}-\vec{r}_{a}(t))\,,\label{eq4}
\end{equation}
and the derivative of trace with respect to $\phi$ reads 
\begin{equation} \label{eq5}
    \frac{\partial T}{\partial \phi}=-c^{2}\sum_{a}\frac{m_{a}}{\phi_{0}}\left[s_{a}\left(1-\frac{1}{2}h^{k}_{k}-\frac{v_{a}^{2}}{2c^{2}}\right)-(s_{a}-s_{a}^{2}-s^{\prime}_{a})\frac{\tilde{\phi}}{\phi_{0}}\right]\delta^{3}(\vec{r}-\vec{r}_{a}(t)).
\end{equation}
Given (\ref{eq4}) and (\ref{eq5}), we can solve (\ref{3.5}) and (\ref{3.4}) upto 1PN order to obtain the leading order static solution for $\tilde \phi$\cite{Hohmann:2015kra},
\begin{equation}
    \tilde{\phi}=\frac{4\xi}{c^{2}}\sum_{a}\frac{m_{a}}{r_{a}}\Big(1-2\frac{s_{a}}{\phi_{0}}\frac{G_{4(0,0)}}{G_{4(1,0)}}\Big)\,,
\end{equation}
and the tensor field  $\tilde h_{\mu\nu}$ within 1PN near zone limit,
\begin{align}
    \begin{split}
        & \tilde{h}_{00}=\frac{4}{c^{2}G_{4(0,0)}}\sum_{a}\frac{m_{a}}{r_{a}}+\mathcal{O}\left(\frac{v}{c}\right)^{4}\,, \\ & \tilde{h}_{ij}=\frac{4v_{i}v_{j}}{c^{4}G_{4(0,0)}}\sum_{a}\frac{m_{a}}{r_{a}}+\mathcal{O}\left(\frac{v}{c}\right)^{6}\,, \\ & \tilde{h}=-\frac{4}{c^{2}G_{4(0,0)}}\sum_{a}\frac{m_{a}}{r_{a}}+\mathcal{O}\left(\frac{v}{c}\right)^{4}.
    \end{split}
\end{align}
Using (\ref{3.3}), we get the following expressions for metric perturbations at leading order
\begin{equation}
    h_{00}=\frac{2}{c^{2}G_{4(0,0)}}\sum_{a}\frac{m_{a}}{r_{a}}+\frac{4\xi}{c^{2}}\frac{G_{4(1,0)}}{G_{4(0,0)}}\sum_{a}\frac{m_{a}}{r_{a}}e^{-m_{s}r_{a}}\left(1-2\frac{s_{a}}{\phi_{0}}\frac{G_{4(0,0)}}{G_{4(1,0)}}\right)+\mathcal{O}\left(\frac{v}{c}\right)^{4}\,,
\end{equation}
\begin{equation}
    h_{ij}=\delta_{ij}\left[\frac{2}{c^{2}G_{4(0,0)}}\sum_{a}\frac{m_{a}}{r_{a}}-\frac{4\xi}{c^{2}}\frac{G_{4(1,0)}}{G_{4(0,0)}}\sum_{a}\frac{m_{a}}{r_{a}}e^{-m_{s}r_{a}}\left(1-2\frac{s_{a}}{\phi_{0}}\frac{G_{4(0,0)}}{G_{4(1,0)}}\right)\right]+\mathcal{O}\left(\frac{v}{c}\right)^{4}\,,
\end{equation}
\begin{equation}
    h=\frac{4}{c^{2}G_{4(0,0)}}\sum_{a}\frac{m_{a}}{r_{a}}-\frac{16\xi}{c^{2}}\frac{G_{4(1,0)}}{G_{4(0,0)}}\sum_{a}\frac{m_{a}}{r_{a}}e^{-m_{s}r_{a}}\left(1-2\frac{s_{a}}{\phi_{0}}\frac{G_{4(0,0)}}{G_{4(1,0)}}\right)+\mathcal{O}\left(\frac{v}{c}\right)^{4}\,.
\end{equation}


 \section{EIH Equations of motion}\label{sec5}
 As mentioned in Sec.~(\ref{sec1}), the Horndeski theory theory violates GWEP due to the fact that mass of the binary $m_{a}$ depends on $\phi$ which in turn is space-time dependent. This dependence leads to the fact that different bodies will respond differently to different fields. They'll have different trajectories. To be more specific the mass is dependent on sensitivity parameters. In \cite{Einstein}, authors suggested a method by which we can see the effects of these sensitivities on the equations of motion and it starts with a Lagrangian which is called Einstein-Infeld-Hoffmann Lagrangian, $L_{EIH}.$
 \begin{align}
     \begin{split} \label{EIH}
         & L_{EIH}=-c^{2}\sum_{a}\int m_{a}(\phi) \frac{d\tau_{a}}{dt} \\& =-c^{2}\sum_{a}m_{a}(\phi)\sqrt{-g_{00}-2g_{0i}\frac{v^{i}_{a}}{c}-g_{ij}\frac{v^{i}_{a}v^{j}_{a}}{c^{2}}}\\& =-\sum_{a}m_{a}c^{2}\Big\{1-\frac{v_{a}^{2}}{2c^{2}}-\sum_{b\neq a}\Big[\frac{1}{c^{2}G_{4(0,0)}}\frac{m_{b}}{r_{ab}}+\frac{2\xi}{c^{2}}\frac{G_{4(1,0)}}{G_{4(0,0)}}\frac{m_{b}}{r_{ab}}\left(1-\frac{2s_{b}}{\phi_{0}}\frac{G_{4(0,0)}}{G_{4(1,0)}}\right)e^{-m_{\tilde{\phi}}r_{ab}}\\ & -\frac{4s_{a}\xi}{c^{2}\phi_{0}}\frac{m_{b}}{r_{ab}}\left(1-\frac{2s_{b}}{\phi_{0}}\frac{G_{4(0,0)}}{G_{4(1,0)}}\right)e^{-m_{\tilde{\phi}} r_{ab}}\Big]+\mathcal{O}(\frac{v}{c})^{4} \Big\}.
         \end{split}
 \end{align}
 Here, $r_{ab}=|\vec{r}_{a}(t)-\vec{r}_{b}(t)|$. The effective gravitational constant can be read off from the above equation and takes the following form, 
 \begin{equation}
     G_{ab}=\left[\frac{1}{G_{4(0,0)}}+2\xi \frac{G_{4(1,0)}}{G_{4(0,0)}}\left(1-\frac{2s_{b}}{\phi_{0}}\frac{G_{4(0,0)}}{G_{4(1,0)}}\right)e^{-m_{\tilde{\phi}}r_{ab}}-\left(1-\frac{2s_{b}}{\phi_{0}}\frac{G_{4(0,0)}}{G_{4(1,0)}}\right)e^{-m_{\tilde{\phi}} r_{ab}}\frac{4s_{a}\xi}{\phi_{0}}\right].
 \end{equation}
 When we substitute the Lagrangian (\ref{EIH}) into the Euler-Lagrange equation we get the $n$-body equations of motion at the leading order
 \begin{equation}
     \vec{a}_{i}=-\sum_{i\neq j}\frac{\mathcal{G}_{ij}m_{j}}{r^{2}_{ij}}\hat{r}_{ij}\,,
 \end{equation}
 with 
 \begin{align}
 \begin{split}
     \mathcal{G}_{ij}=\frac{1}{G_{4(0,0)}}&\Big\{1+(1+m_{\tilde{\phi}}r_{ab})e^{-m_{\tilde{\phi}}r_{ab}} \Big[2\xi G_{4(1,0)}\left(1-\frac{2s_{b}}{\phi_{0}}\frac{G_{4(0,0)}}{G_{4(1,0)}}\right)\\&-\frac{4s_{a}\xi G_{4(0,0)}}{\phi_{0}}\left(1-\frac{2s_{b}}{\phi_{0}}\frac{G_{4(0,0)}}{G_{4(1,0)}}\right)\Big] \Big\}. \label{effectivegg}
 \end{split}
 \end{align}
Here, $\vec{a}_{i}\equiv \frac{d^{2}\vec{r}_{i}}{dt^{2}}$ is the acceleration of the $i^{th}$ object, $\hat{r}_{ij}$ is the unit direction vector from the $j^{th}$ particle to the $i^{th}$ particle. The terms having exponents cause the inverse square law to not hold in this case. However, $m_{s}$ having dimension of inverse length which is of the order of cosmological scales and $r_{ab}$ being small compared to $m_{s}$ makes $m_{s}r_{ij}\ll 1$. Under these approximations  $e^{-m_{\tilde{\phi}}r_{ab}}\rightarrow 1$ and 
  \begin{equation} \label{eq8}
     \mathcal{G}_{ab}=\frac{1}{G_{4(0,0)}}\left\{1+2\xi G_{4(1,0)}\left(1-\frac{2s_{b}}{\phi_{0}}\frac{G_{4(0,0)}}{G_{4(1,0)}}\right)-\frac{4s_{a}\xi G_{4(0,0)}}{\phi_{0}}\left(1-\frac{2s_{b}}{\phi_{0}}\frac{G_{4(0,0)}}{G_{4(1,0)}}\right) \right\}.
 \end{equation}
 Note that although the inverse square law form is restored the GWEP violation still happens. However, due to this inverse square law form we can now apply the usual orbital dynamical equations to our problem. We can calculate the orbital decay rate $\dot{P}$ given by $\frac{\dot{P}}{P}=-\frac{3}{2}\frac{\dot{E}}{E}$, P in turn satisfying the usual Kepler's third law 
 \begin{equation}
     a^{3}\left(\frac{2\pi}{P}\right)^{2}=\mathcal{G}_{12}M\,,
 \end{equation}
 and $E=-\frac{\mathcal{G}_{12}M\mu}{2a}$, where a is the semi-major axis, $\mathcal{G}_{12}$ is the effective gravitational constant. $M$ is the total mass of this composite system. These relations will give us the energy loss of the system. But before that we need the energy-momentum tensor to complete the analysis which will be done in the next section.
 
\section{Stress Tensor}\label{5}
In this section, we want to introduce the tools to calculate the radiations from the system. The system radiates away energy and also suffers a loss in angular momentum. The emission of these radiations include a scalar part and a gravitational part, and the radiations are monopole, dipole, octupole and dipole-octupole type. These will give us the necessary orbital decay rate and, eventually, the change in the orbital period. To find this, we need to first calculate the stress tensor of this system. There are many methods to calculate this (pseudo) stress-tensor. But, all these results give different stress tensors but lead to identical flux expressions \cite{Petrov,Saffer:2017ywl}. In \cite{Dyadina:2018ryl}, authors used Noether's procedure to derive a non-symmetric stress tensor. But it poses no problem for calculating the energy flux. However, difficulty arises during the computation of angular momentum, for which we have to use a symmetric pseudo-tensor. In this paper, we have found a  symmetric pseudo tensor using the method mentioned in \cite{Isaacson:1968hbi,Isaacson:1968zza} which we describe below.
\par 
The approach works in the short wavelength limit, where  $\lambda \ll \frac{1}{\sqrt{R}}$, where $R$ are the values of the background Riemann tensor components. Our background is flat, hence $R=0$ and thus the condition is always satisfied and now the task remains is to take average of the quantities which requires the following rules \cite{Hou:2017cjy,MTW}:
\begin{itemize}
    \item $ \langle \partial_{\alpha}(\tilde{h}_{\beta \gamma}\partial_{\delta}\tilde{h})\rangle=0\,,$
    \item $\langle \tilde{h}\partial_{\alpha}\partial_{\beta}\tilde{h}_{\gamma \delta}\rangle=-\langle\partial_{\alpha}\tilde{h}\partial_{\beta}\tilde{h}_{\gamma \delta} \rangle $
\end{itemize}
Same rules are applied for the terms involving scalar field as well.
\par 
In \cite{Hou:2017cjy}, this method has been used to compute a symmetric stress-tensor in vacuum using transverse-traceless gauge $(\partial_{\alpha}\tilde{h}^{\alpha \beta}=0 , \tilde{h}=0).$ It is given by,
\begin{equation} \label{stress}
    \langle T_{\mu \nu} \rangle=\Big \langle \frac{c^{4}}{32\pi}G_{4(0,0)}\partial_{\mu}\tilde{h}^{,\rho \sigma}\partial_{\nu}\tilde{h}_{\rho \sigma}+ \frac{2 \xi\,G_{4(0,0)}}{G_{4(1,0)}}\partial_{\mu}\tilde{\phi}\partial_{\nu}\tilde{\phi}+m_{\tilde{\phi}}^{2}G_{4(1,0)}\tilde{\phi}\,\tilde{h}_{\mu \nu}\Big \rangle\,.
\end{equation}

In the limit that $G_{4}=\frac{1}{G_{N}}$ and the remaining $G_{i}$'s vanish we get the effective stress tensor for GR. We now have everything to proceed to calculate the fluxes save for the solution of $\phi$ which will be done in the next section.

\section{Radiations from the Binary Pulsars}\label{sec6}
The previous section gives us the energy-momentum tensor. However, we are yet to find the solution of $\tilde{\phi}$ so that we can calculate the $\partial_{\mu}\tilde{\phi}\partial_{\nu}\tilde{\phi}$ terms in addition to the $\tilde{h}$ terms. To compute the energy-momentum tensor we need the solutions to the following equations
   \begin{equation}
   \Box \tilde{\phi}-m_{s}^{2}\tilde{\phi}=\frac{16 \pi}{c^{4}}\xi\left[T^{(1)}-\frac{2G_{4(0,0)}}{G_{4(1,0)}}\frac{\partial T}{\partial \phi}^{(1)}\right]\,,
\end{equation} 
\begin{equation}
    \Box \tilde{h}_{\mu \nu}=-\frac{16 \pi}{c^{4}G_{4(0,0)}}T_{\mu \nu}^{(1)}\,.
\end{equation}
\subsection{Solution to the tensor part}
The solution to 
\begin{equation}
    \Box \tilde{h}_{\mu \nu}=-\frac{16 \pi}{c^{4}G_{4(0,0)}}T_{\mu \nu}^{(1)}
\end{equation}
is obtained by Green's function method and takes the following form \cite{Poisson},
\begin{equation}
    \tilde{h}_{\mu \nu}=\frac{4}{c^{4}G_{4(0,0)}}\int d^{3}r \frac{T^{(1)}_{\mu \nu}(t-|\vec{r}-\vec{r}^{\prime}|/c)}{|\vec{r}-\vec{r}^{\prime}|}\,.
\end{equation}
Since we want to find the solution at the far limit then $|\vec{r}^{\prime}|\ll |\vec{r}|$, $\vec{r}^{\prime}$ being the source point and $\vec{r}^{\prime}$ the point where we want to find the field to. Taking this into account and expanding the denominator we get \cite{Poisson},
\begin{equation}
    \tilde{h}_{\mu \nu}=\frac{4}{r c^{4}G_{4(0,0)}}\sum_{l=0}^{\infty}\frac{1}{c^{l}l!}\frac{\partial^{l}}{\partial t^{l}}\int d^{3}r T^{(1)}_{\mu \nu}(t-r/c,\vec{r}^{\prime})(\hat{n}.\vec{r}^{\prime})^{l}
\end{equation}
where, $\hat{n}=\frac{\vec{r}}{r}$ being the unit vector. Upto leading order($l=0$) we can use the conservation equation for the $T^{(1)}_{\mu \nu}$ tensor to rewrite it as
\begin{equation}
    \tilde{h}_{ij}=\frac{4}{r c^{4}G_{4(0,0)}}\int d^{3}r^{\prime}T^{(1)}_{ij}(t-r/c,\vec{r}^{\prime})\,,
\end{equation}
and eventually we get the following,
\begin{equation} \label{eq2}
    \tilde{h}_{ij}=\frac{2}{r c^{6}G_{4(0,0)}}\frac{\partial^{2}}{\partial t^{2}}\int d^{3}r^{\prime}T^{(1)}_{00}(t-r/c,\vec{r}^{\prime})r^{\prime}_{i}r^{\prime}_{j}\,.
\end{equation}
The above relation shows that upto leading order the tensorial radiation is dominated by quadrupole radiation, and that the monopole and dipole radiation are not present. This is exactly like GR. This happens because the graviton is spin-2 as well as massless. At the leading order PN, $T^{(1)}_{00}$ which is the sum total energy density of matter fields, is negligible. So 
\begin{equation}
    T^{(1)}_{00}=\sum_{a}m_{a}c^{2}\delta(\vec{r}-\vec{r}_{a}(t)).
\end{equation}
Substituting this in (\ref{eq2}) we get
\begin{equation}
    \tilde{h}_{ij}=\frac{2}{rc^{4}G_{4(0,0)}}\frac{\partial^{2}}{\partial t^{2}}M_{ij}|_{ret}
\end{equation}
where, \begin{align} \label{quad} M_{ij}=\sum_{a}m^{a}r^{a}_{i}(t)r^{a}_{j}(t)\end{align}
is the mass quadrupole moment \cite{Poisson}. To find the flux for this tensor radiation part we first find out the  transverse-traceless (TT) part of $h_{ij}$ by using the projection operator $\Lambda_{ij,kl}$, called the Lambda tensor. The TT part is $h_{ij,TT}=\Lambda_{ij,kl}h^{kl}=\Lambda_{ij,kl}\tilde{h}^{kl}$ \cite{Poisson}. 
\subsection{Solution to the scalar part}
As mentioned earlier we need to find the solution to the scalar equation also. From (\ref{3.5}) we know that it takes the following form,
\begin{equation}
    \Box \tilde{\phi}-m_{s}^{2}\tilde{\phi}=\frac{16 \pi}{c^{4}}\xi\left[T^{(1)}-\frac{2G_{4(0,0)}}{G_{4(1,0)}}\frac{\partial T}{\partial \phi}^{(1)}\right]=\frac{16 \pi}{c^{4}}\xi \tilde{T}^{(1)}
\end{equation}
where, $\tilde{T}^{(1)}=\left[T^{(1)}-\frac{2G_{4(0,0)}}{G_{4(1,0)}}\frac{\partial T}{\partial \phi}^{(1)}\right]$. Again just like the tensorial part we can find the solution of the scalar equation using Greens function method \cite{Alsing:2011er}. For the sake of simplifying the calculation we assume $G_{2(2,0)}=0$ to make the equation massless. The usual Green's function method works,
\begin{equation} 
    \Box G(x,x^{\prime})=-4\pi \delta^{4}(x-x^{\prime})
\end{equation}
and the solution is  \cite{Alsing:2011er}
\begin{equation} \label{phisol}
    \tilde{\phi}=\frac{4\,\xi}{r\,c^{2}}\sum_{l=0}^{\infty}\frac{1}{c^{l}l!}\frac{\partial^{l}}{\partial t^{l}}\int d^{3}r^{\prime} (\hat{n}.\vec{r}^{\prime})^{l}\tilde{T}^{(1)}.
\end{equation}
Again the integration region is similar to that of the tensorial part, i.e., we find the field at a far point such that $|\vec{r}^{\prime}|\ll |\vec{r}|$. Substituting the source term $\tilde{T}^{(1)}$ from the expressions from (3.9) and after doing the algebra we arrive at the following expression \cite{Alsing:2011er},
\begin{equation} \label{scalar1}
    \tilde{\phi}=\frac{4\,\xi}{r\,c^{2}}\sum_{l=0}^{\infty}\frac{1}{c^{l}l!}n_{i_{1}}n_{i_{1}}n_{i_{2}}n_{i_{3}}...n_{i_{l}}\partial^{l}_{t}\mathcal{M}^{L}_{l}
\end{equation}
where 
\begin{equation}
    \mathcal{M}^{L}_{l}=\mathcal{M}^{i_{1}i_{2}i_{3}....i_{l}}_{l}(t,r,z)=\sum_{a}M_{a}(t-r/c)r^{L}_{a}(t-r/c)\,,
\end{equation}
and 
\begin{align}
\begin{split} \label{Quad1}
 M_{a}(t)=m_{a} & \Big[1-2\frac{G_{4(0,0)}}{G_{4(1,0)}}\frac{s_{a}}{\phi_{0}}-\frac{v_{a}^{2}}{2c^{2}}\left(1-2\frac{G_{4(0,0)}}{G_{4(1,0)}}\frac{s_{a}}{\phi_{0}}\right)-3\sum_{b\neq a}\frac{m_{b}}{r_{ab}(t)c^{2}G_{4(0,0)}}\left(1-2\frac{G_{4(0,0)}}{G_{4(1,0)}}\frac{s_{a}}{\phi_{0}}\right)+\\ & \frac{6\,G_{4(1,0)}\xi}{c^{2}G_{4(0,0)}}\sum_{b\neq a}\frac{m_{b}}{r_{ab}(t)}\left(1-2\frac{G_{4(0,0)}}{G_{4(1,0)}}\frac{s_{a}}{\phi_{0}}\right)-\sum_{b\neq a}\frac{m_{b}}{r_{ab}(t)c^{2}}\left(1-2\frac{G_{4(0,0)}}{G_{4(1,0)}}\frac{s_{a}}{\phi_{0}}\right)\\ &\left(\frac{8\xi s_{a}}{\phi_{0}}-\frac{8}{\phi_{0}}\frac{2G_{4(0,0)}}{G_{4(1,0)}}(s_{a}-s^{2}_{a}+s_{a})\xi \right) \Big].
 \end{split}
\end{align}

\section{Calculation of fluxes}\label{sec7}
We now have the solution for the scalar part of the equation (\ref{scalar1}) and also the tensorial part (\ref{stress}). Then the task remains is to put them into the of the pseudo-tensor. Also, we consider a system of binaries with masses $m_{1}$ and $m_{2}.$ The reduced mass ($\mu$) and the total mass ($M$) of the system is defined as  $\mu=\frac{m_{1}m_{2}}{m_{1}+m_{2}}$ and $M=m_{1}+m_{2}$. We parametrized the orbit by polar coordinates $(r,\theta)$ and set the origin at the centre of mass of the system. Then we fix our frame at the centre of mass. While doing this we choose the following parametrization \cite{Maggiore:2007ulw},
\begin{equation}
    x=r\cos\theta,y=r\sin\theta,z=0
\end{equation}
with, 
\begin{equation}
    r=\frac{a(1-e^{2})}{1-e\cos\theta}
\end{equation}
where, $e$ is the eccentricity and $a$ is the semi-major axis of the binary orbit. Time derivative of $\theta$ can be computed by noting that, angular momentum ($L$) is give by,  $L=\mu\, r^{2}\,\dot{\theta}.$ Then utilizing this we get,
\begin{equation}
    \dot{\theta}=\sqrt{\frac{\mathcal{G}_{12}\,M}{G_{4(0,0)}a^{3}}}(1-e^{2})^{-3/2}(1+e\cos\theta)^{2}
\end{equation}
where, $\mathcal{G}_{12}$ is defined in (\ref{eq8}). Also, henceforth we will label the two bodies that the binary system is composed of by $1$ and $2.$
\subsection{Energy loss for the binaries} \label{Energyflux}
The component of the energy momentum pseudo-tensor contributing to the energy flux of the system is $T_{0r}$  and we can calculate the energy-flux from the following integral \cite{Maggiore:2007ulw},
\begin{equation}
    \langle \dot{E} \rangle=-c\, r^{2}\int \langle T _{0r}\rangle\, d\Omega\,.
\end{equation}
where, $\langle T_{0r} \rangle$ is defined in (\ref{stress}). At this point for calculation ease we have set $G_{2(2,0)}$ to zero which kills the third term in in (\ref{stress}). Also it helps to perform the orbital averaging needed to obtain expressions for various fluxes. Contribution to the energy flux due to the tensor part of the pseudo-tensor takes the following form,
\begin{align}
    \begin{split}
        & \langle \dot{E}_{T}\rangle=\frac{c^{5}r^{2}}{16\pi}\int d\Omega\, \Big \langle \frac{G_{4(0,0)}}{2}\partial_{0}\tilde{h}^{\mu \nu} \partial_{r}\tilde{h}_{\mu \nu} \Big \rangle=-\frac{c^{5}r^{2}G_{4(0,0)}}{32\pi}\int d\Omega \, \langle \partial_{0}\tilde{h}^{\mu \nu}\partial_{0}\tilde{h}_{\mu \nu}  \rangle.
    \end{split}
\end{align}
Then putting the solution for $\tilde{h}_{ij}$ from (6.9) we get
\begin{equation}
    \langle \dot{E}_{T}\rangle=-\frac{1}{5\,c^{5}G_{4(0,0)}}\,\Big \langle\dddot{M}^{kl}\dddot{M}^{kl}-\frac{1}{3}(\dddot{M}^{kk})^{2}\Big\rangle.
\end{equation}
Upon inserting the value of $M_{ij}$ from (\ref{quad}) and performing an orbital averaging over the orbital period of the binary\footnote{For more details on orbital averaging the interested reader can refer to \cite{Maggiore:2007ulw}.} we get the following, 
\begin{align}
\begin{split}
 \langle \dot{E}_{T}\rangle=   -\frac{32\mu^{2}M^{3} \mathcal{G}_{12
 }^3}{5\,c^{5}G_{4(0,0)}\,a^{5}}\frac{1}{(1-e^{2})^{7/2}}\left(1+\frac{73}{24}e^{2}+\frac{37}{96}e^{4}\right) \,,
    \end{split}
\end{align}
where $\mathcal{G}_{ab}$ is defined in (\ref{eq8}).\par
Now we can focus on the scalar energy flux. It takes the following form,
\begin{align}
    \begin{split}
       \langle \dot{E}_{\phi}\rangle= & \frac{c^{5}r^{2}}{16\pi}\int d\Omega \, \, \Big \langle \frac{G_{4(1,0)}}{2G_{4(0,0)}\xi}\partial_{0}\tilde{\phi}\partial_{r}\tilde{\phi}\Big \rangle=-\frac{c^{5}r^{2}G_{4(1,0)}}{32\pi G_{4(0,0)}}\int d\Omega \langle \partial_{0}\tilde{\phi}\partial_{r}\tilde{\phi} \rangle
    \end{split}
\end{align}
Using the solution for $\tilde{\phi}$ mentioned in (\ref{phisol}) and neglecting the terms $\mathcal{O}(1/r^{2})$ while taking the derivatives we get
\begin{align}
\begin{split} \label{eephi}
 \langle \dot{E_{\phi}}\rangle=& -\frac{2\,c^5\, G_{4(1,0)}\,\xi}{G_{4(0,0)}}\Big\langle \frac{1}{c^{6}}\dot{\mathcal{M}}_{0}\dot{\mathcal{M}}_{0}+\frac{1}{3c^{8}}\ddot{\mathcal{M}}_{1}^{k}\ddot{\mathcal{M}}_{1}^{k}+\frac{1}{3c^{8}}\dot{\mathcal{M}}_{0}\dddot{\mathcal{M}}^{kk}_{2}\\ & +\frac{1}{30c^{10}}\dddot{\mathcal{M}}_{2}^{kl}\dddot{\mathcal{M}}_{2}^{kl}+\frac{1}{60c^{10}}\dddot{\mathcal{M}}_{2}^{kk}\dddot{\mathcal{M}}_{2}^{ll}+\frac{2}{15c^{10}}\ddot{\mathcal{M}}_{1}^{k}\ddddot{\mathcal{M}}_{3}^{kll}\Big\rangle,
\end{split}
\end{align}
where, 
\begin{align}
    \begin{split}
  &    \langle\dot{\mathcal{M}}_{0}\dot{\mathcal{M}}_{0}\rangle=\frac{2\,A^{2}\mathcal{G}_{12} \mu^{2} M^{3}}{a^{5}(1-e^{2})^{7/2}} e^{2}\left(1+\frac{e^{2}}{4}\right),\, \langle \dot{\mathcal{M}}_{0}\dddot{\mathcal{M}}^{kk}_{2}\rangle=\frac{7\tilde{\alpha}^{2}A\tilde{A}\mu^{2}M}{2a^{5}(1-e^{2})^{7/2}}e(1+\frac{e^{2}}{7}),\,
        \\ & \langle \dddot{\mathcal{M}}^{kk}_{2}\dddot{\mathcal{M}}^{ll}_{2}\rangle=\frac{\tilde{\alpha}^{3}\mu^{2}\tilde{A}^{2}}{a^{5}(1-e^{2})^{7/2}}\left(32+84e^{2}+\frac{57}{8}e^{4}\right), \langle \dddot{\mathcal{M}}^{kl}_{2}\dddot{\mathcal{M}}^{kl}_{2}\rangle=\frac{\tilde{\alpha}^{3}\mu^{2}\tilde{A}^{2}}{a^{5}(1-e^{2})^{7/2}}(64+170e^{2}+21e^{4}),
        \\ & \langle \ddot{\mathcal{M}}^{k}_{1}\ddddot{\mathcal{M}}^{kll}_{3}\rangle=\frac{\tilde{\alpha}^{3}\mu^{2}}{a^{5}(1-e^{2})^{7/2}}\Big[A_{1}A_{d}\Big(1-\frac{132}{8}e^{2}-\frac{51}{8}e^{4}\Big)-A_{1}\tilde{A}_{d}\Big(1-\frac{2192}{64}e^{2}-\frac{3368}{64}e^{4}-\frac{279}{64}e^{6}\Big)\Big],\\&
    \langle \ddot{\mathcal{M}}^{k}_{1}\ddot{\mathcal{M}}^{k}_{1}\rangle=\frac{\tilde{\alpha}^{2}}{a^{6}(1-e^{2})^{9/2}}\Big[A^{2}_{d}a^{2}(1-e^{2})^{2}\left(1+\frac{e^{2}}{2}\right)+\frac{\tilde{A}^{2}_{d}}{64}(64+608e^{2}+552e^{4}+36e^{6})+\\ & 2A_{d}\tilde{A}_{d}\left(1+3e^{2}+\frac{3}{8}e^{4}\right)\Big].
    \end{split} \label{scalarmult}
\end{align}
Here we have used the multipole expansion mentioned in (\ref{eephi}). The first term of (\ref{eephi}) is a monopole contribution. The second and third terms of (\ref{eephi}) are dipole-dipole and monopole-dipole cross term respectively. The fourth and fifth corresponds to quadrupole radiations and the last term is again a cross term between dipole-octupole combination. For more details interested readers are referred to \cite{Zhang:2017srh}. Here we have used (\ref{scalar1}) and (\ref{Quad1}). The details of the computation is given in Appendix~(\ref{A}). Please note that we have truncated our expression in (\ref{eephi}) upto the mixed dipole-octupole term $\ddot{\mathcal{M}}_{2}^{k}\ddddot{\mathcal{M}}_{3}^{kll}.$ \par
Then finally putting all these together we get,
\begin{align}
\begin{split} \label{Edot}
 \langle \dot{E}\rangle=& -\frac{32\mu^{2}M^{3}\mathcal{G}_{12}^{3}}{5c^{5}G_{4(0,0)}a^{5}} \frac{1}{(1-e^{2})^{7/2}}\left(1+\frac{73}{24}e^{2}+\frac{37}{96}e^{4}\right)-\frac{2G_{4(1,0)}\xi}{G_{4(0,0)}}\Big[ \frac{2\,A^{2}\mathcal{G}_{12} \mu^{2} M^{3}}{c\, a^{5}(1-e^{2})^{7/2}} e^{2}\left(1+\frac{e^{2}}{4}\right)\\& +\frac{\tilde{\alpha}^{2}}{3c^{3}\,a^{6}(1-e^{2})^{9/2}}\Big\{A^{2}_{d}a^{2}(1-e^{2})^{2} \left(1+\frac{e^{2}}{2}\right)+ \frac{\tilde{A}^{2}_{d}}{64}(64+608e^{2}+552e^{4}+36e^{6})+\\&  2A_{d}\tilde{A}_{d}\left(1+3e^{2}+\frac{3}{8}e^{4}\right)\Big\}+\frac{7\,\tilde{\alpha}^{2}\,A\,\tilde{A}\,\mu^{2}M}{6c^{3}a^{5}(1-e^{2})^{7/2}}e\Big(1+\frac{e^{2}}{7}\Big) +\frac{1}{60c^{5}}\frac{\tilde{\alpha}^{3}\mu^{2}\tilde{A}^{2}}{a^{5}(1-e^{2})^{7/2}}\\ &(160 +424e^{2}+\frac{393}{8}e^{4}) +\frac{2}{15c^{5}}\frac{\tilde{\alpha}^{3}\mu^{2}}{a^{5}(1-e^{2})^{7/2}}\Big\{A_{1}A_{d}\Big(1-\frac{132}{8}e^{2}-\frac{51}{8}e^{4}\Big)-\\ &A_{1}\tilde{A}_{d}\Big(1-\frac{2192}{64}e^{2}-\frac{3368}{64}e^{4}-\frac{279}{64}e^{6}\Big)\Big\}  \Big],
  \end{split}
\end{align}
where,
\begin{equation}
    A=-\frac{1}{c^{2}}\Big[\Big(\frac{3}{G_{4(0,0)}}-\frac{6G_{4(1,0)}}{G_{4(0,0)}}+\gamma_{1}\Big)\beta_{2}+\Big(\frac{3}{G_{4(0,0)}}-\frac{6G_{4(1,0)}}{G_{4(0,0)}}+\gamma_{2}\Big)\beta_{1} \Big]\,,\nonumber
\end{equation}
\begin{equation}
    A_{d}=\mu (\alpha_{1}-\alpha_{2}),
    \tilde{A}=1-\alpha^{\prime}\,,A_{1}=\frac{m^{2}_{2}-m^{2}_{1}}{m^{2}_{1}m^{2}_{2}}-\Big( \frac{\alpha_{1}}{m_{1}^{2}}-\frac{\alpha_{2}}{m_{2}^{2}}\Big)\,,\nonumber
\end{equation}
\begin{equation}
    \tilde{A}_{d}=\frac{\mu}{c^{2}}\Big[ \Big(\frac{3}{G_{4(0,0)}}-\frac{6G_{4(1,0)}}{G_{4(0,0)}}+\gamma_{1}\Big)m_{2}\beta_{2}+\Big(\frac{3}{G_{4(0,0)}}-\frac{6G_{4(1,0)}}{G_{4(0,0)}}+\gamma_{2}\Big)m_{1}\beta_{1}\Big]\,,\label{definition1}
\end{equation}
\begin{align}
\begin{split}
 & \alpha_{1,2}=2\frac{G_{4(0,0)}}{G_{4(1,0)}}\frac{s_{1,2}}{\phi_{0}}, \beta_{1,2}=\Big(1-\frac{2G_{4(0,0)}}{G_{4(1,0)}}\frac{s_{1,2}}{\phi_{0}}\Big),\\& \gamma_{1,2}= \left(\frac{8\xi s_{1,2}}{\phi_{0}}-\frac{8}{\phi_{0}}\frac{2G_{4(0,0)}}{G_{4(1,0)}}(s_{1,2}-s^{2}_{1,2}+s_{1,2})\xi \right)\,,\label{definition}
  \end{split}
\end{align}
\begin{equation}
    \alpha^{\prime}=\frac{\alpha_{1}m_{2}+\alpha_{2}m_{1}}{m_{1}+m_{2}},\,\, \tilde{\alpha}=\sqrt{\frac{\mathcal{G}_{12} M}{a^{3}}}.\label{definition2}
\end{equation}
One can arrive at a simple expression for circular orbit by setting $e=0.$ The expression in (\ref{Edot}) the becomes,
\begin{align}
\begin{split} \label{Ecdot}
 \langle \dot{E}\rangle=& -\frac{32\mu^{2}M^{3}\mathcal{G}_{12}^{3}}{5c^{5}G_{4(0,0)}a^{5}}-\frac{2G_{4(1,0)}\xi}{G_{4(0,0)}}\Big[\frac{\tilde{\alpha}^{2}}{3c^{3}a^{6}}\Big\{A^{2}_{d}a^{2}+ \tilde{A}_{d}(\tilde{A}_{d}+ 2\, A_{d})\Big\} +\frac{2\,\tilde{\alpha}^{3}\mu^{2}}{15\, c^5\,a^{5}}\Big\{20\,\tilde{A}^{2} +A_{1}\,A_{d}-A_{1}\tilde{A}_{d}\Big\}  \Big],
  \end{split}
\end{align}
$a$ becomes the radius of the circular orbit. We now make few comments. \begin{itemize}
    \item 
Note that the first term in (\ref{Ecdot}) comes from the tensorial part of the radiation and has the same form as GR. The correction  due to Horndeski couplings are present inside the effective gravitational constant $\mathcal{G}_{12}$ defined in (\ref{eq8}). The next three terms are due to the scalar radiation. In the circular limit the monopole contribution i.e $\langle\dot{\mathcal{M}}_{0}\dot{\mathcal{M}}_{0}\rangle$  vanishes  as evident from (\ref{scalarmult}). 
\item From (\ref{definition}) and (\ref{definition2}), it is evident that $A_d$  and $\tilde \alpha$ are independent of $c.$ Hence the leading contribution to the total energy flux comes from the scalar part and comes at $\mathcal{O}(\frac{1}{c^3})$. \footnote{It is well-known in the literature that the dipole radiation, due to the presence of a scalar field, gives a dominant contribution to the energy flux expression compared to the tensor part. This is true even for other theories of gravity as long as a scalar field is present in theory, e.g. interested readers are referred to \cite{tanja} for similar computation for scalar-Gauss Bonnet theory.} This is also observed in \cite{Zhang:2017srh,Dyadina:2018ryl}. Note that, in \cite{Zhang:2017srh}, authors have considered scalar-tensor theory, a subclass of Horndeski theory, in the context of scalar radiation, and the results serve as a consistency check for our results. It is solely coming from the dipole term $ \langle \ddot{\mathcal{M}}^{k}_{1}\ddot{\mathcal{M}}^{k}_{1}\rangle.$ The contribution from monopole-dipole term vanishes in this circular limit. \item Finally, the last three terms of (\ref{Ecdot}) correspond to quadrupole and mixed dipole-octupole terms  \footnote{Note that our results for $e=0$ case cannot be directly compared with results of \cite{Dyadina:2018ryl}  as our computation is valid only when the mass of the scalar field is set to zero. However, a qualitative similarity with our results exists.}.
\end{itemize}
    
\subsection{Angular Momentum loss rate for the binaries} \label{angularmomentumflux}
Just like in the Sec.~(\ref{Energyflux}) where we calculated the energy flux for the binaries using the components of the energy-momentum tensor defined in (\ref{stress}), the part of the tensor which contributes to the angular momentum loss rate is given by $T_{ij}$ component. Then we can calculate the angular momentum flux from the following integral \cite{Maggiore:2007ulw},
\begin{equation}
    \langle \dot{L}^{i}\rangle=-c\, r^{2}\epsilon^{ijk}\int d\Omega\, \langle T^{jk}\rangle\,.
\end{equation}
We can repeat the same step by step calculation from the previous section to obtain the integrals for calculating the tensor part and the scalar part of $T^{jk}$. The integral for the tensor part takes the following form,
\begin{align}
    \begin{split} \label{angpart1}
        & \langle \dot{L}_{T}\rangle=-\epsilon^{ijk}\frac{c^{4}r^{2}G_{4(0,0)}}{32\pi}\int d\Omega\, \langle \partial_{0}\tilde{h}^{\mu \nu}\,x^{j}\partial^{k}\tilde{h}_{\mu \nu}-2\partial_{0}\tilde{h}_{ak}\tilde{h}_{aj} \rangle\,.
    \end{split}
\end{align}
Then putting the solution for $\tilde{h}_{ij}$ mentioned in (\ref{eq2}) and performing the orbital averaging we get,
\begin{align}
\begin{split}
\langle \dot{L}_{T}\rangle  = &-\frac{32\mu^{2}( \mathcal{G}_{12}\,M)^{5/2}}{5c^{5}G_{4(0,0)}\,a^{7/2}} \frac{1}{(1-e^{2})^{2}}\left(1+\frac{7}{8}e^{2}\right).
    \end{split}
\end{align}

where $\mathcal{G}_{12}$ is defined in (\ref{eq8}).  \par 
For the scalar part, after using  the solution for $\tilde{\phi}$ in (\ref{scalar1}) we get,
\begin{align}
\begin{split} 
  &  \langle \dot{L}^{i}_{\phi} \rangle=c\,r^{2}\epsilon^{ikm}\int d\Omega \, \Big \langle \frac{G_{4(1,0)}}{2\,G_{4(0,0)}}\frac{c^{4}}{16\pi}\frac{1}{\xi}\partial_{0}\tilde{\phi}\ x^{k}\ \partial^{m}\tilde{\phi}\Big \rangle\\ & = -\frac{2\,c^{5}G_{4(1,0)}\xi}{3 G_{4(0,0)}}\epsilon^{ikm}\Big\langle \frac{1}{c^{6}}\dot{\mathcal{M}}^{k}_{1}\ddot{\mathcal{M}}^{m}_{1}+\frac{1}{5\,c^{8}}\ddot{\mathcal{M}}^{k\,p}_{2}\dddot{\mathcal{M}}^{m\,p}_{2}+\frac{1}{5\,c^{10}}\dddot{\mathcal{M}}^{llk}_{3}\ddot{\mathcal{M}}^{m}_{1} \Big \rangle\,,\label{angpart2}
    \end{split}
\end{align}

where, 
\begin{align}
    \begin{split}
   &     \epsilon^{ikm}\langle \dot{\mathcal{M}}^{k}_{1}\ddot{\mathcal{M}}^{m}_{1}\rangle = \frac{\mathcal{G}_{12}M}{a^{7/2}(1-e^{2})^{9/2}}\Big\{A^{2}_{d}(1-e^{2})^{2}+\tilde{A}^{2}_{d}(1+3e^{2}+\frac{3}{8}e^{4})+\frac{1}{2}A_{d}\tilde{A}_{d}(1-e^{2})(3+\frac{5}{4}e^{2})\Big\}\,, \label{dipoleang}
   \end{split}
\end{align}

\begin{align}
    \begin{split} \label{quadang}
        & \epsilon^{ikm}\langle \ddot{\mathcal{M}}^{kp}_{2}\dddot{\mathcal{M}}^{mp}_{2}\rangle=\frac{8}{15}\frac{\mu^{2}\mathcal{G}_{12}^{3} M^{3}}{a^{7/2}(1-e^{2})}\left(1+\frac{7}{8}e^{2}\right)\,,
    \end{split}
\end{align}
\begin{align}
    \begin{split}
        & \epsilon^{ikm}\langle\dddot{\mathcal{M}}^{llk}_{3}\ddot{\mathcal{M}}^{m}_{1}\rangle=\frac{\tilde{\alpha}^{5/2}A_{1}A_{d}}{10\, a^{7/2}(1-e^{2})^{2}}\left(1-\frac{7}{2}e^{2}\right)-\frac{\tilde{\alpha}^{5/2}A_{1}\tilde{A}_{d}}{10\,{a^{15/2}(1-e^{2})^{6}}}\left(3-\frac{79}{2}e^{2}-\frac{81}{4}e^{4}\right)\,.
    \end{split}
\end{align}
Like in Sec.~(\ref{Energyflux}) we can identify the first, second and the third term in the second line of  (\ref{angpart2}) as dipole, quadrupole and mixed dipole-octupole contributions respectively. Finally summing up the tensor and  scalar contributions  from (\ref{angpart1}) and (\ref{angpart2}) we get the full expression for the angular momentum flux.

\begin{align}
    \begin{split} \label{Ldot}
   & \langle \dot{L}^{i}\rangle=  -\frac{32\mu^{2}( \mathcal{G}_{12}\,M)^{5/2}}{5c^{5}G_{4(0,0)}\,a^{7/2}} \frac{1}{(1-e^{2})^{2}}\left(1+\frac{7}{8}e^{2}\right)-\frac{2G_{4(1,0)}\xi}{3 G_{4(0,0)}}\Big[ \frac{1}{c}\frac{\tilde{\alpha}^{2}}{a^{6}(1-e^{2})^{9/2}}\Big\{A^{2}_{d}(1-e^{2})^{2}\\ &+\tilde{A}^{2}_{d}(1+3e^{2}+\frac{3}{8}e^{4})+\frac{1}{2}A_{d}\tilde{A}_{d}(1-e^{2})(3+\frac{5}{4}e^{2})\Big\}+\frac{8}{75c^{3}}\frac{\mu^{2}\mathcal{G}_{12}^{3}\, M^{3}}{a^{7/2}(1-e^{2})}\left(1+\frac{7}{8}e^{2}\right)\\ & +\frac{\tilde{\alpha}^{5/2}A_{1}}{50c^{5}10\, a^{7/2}(1-e^{2})^{2}}\Big\{ A_{d}\left(1-\frac{7}{2}e^{2}\right)-\frac{\tilde{A}_{d}}{{a^{4}(1-e^{2})^{4}}}\left(3-\frac{79}{2}e^{2}-\frac{81}{4}e^{4}\right)\Big\}\Big]
    \end{split}
\end{align}

where, $A_d, \tilde A_d, A_1$, and $\tilde \alpha$ are defined in Eqs.~(\ref{definition1}), (\ref{definition}) and (\ref{definition2}).
Before ending this section, we now make few comments. 
\begin{itemize}
\item The first term in (\ref{Ldot}) comes from the tensorial part of the radiation and has the same form as GR. The correction  due to Horndeski couplings are present inside the effective gravitational constant $\mathcal{G}_{12}$ defined in (\ref{eq8}).
\item First of all, note that no monopole contribution exists to the angular momentum flux. As the scalar field is of spin-zero, the monopole part of the radiation carries energy flux as in (\ref{Edot}) but not angular momentum flux. 
\item Also note that $\tilde \alpha$ and $A_d$ are independent of $c.$ Just like what we have done in Sec.~(\ref{Energyflux}), we can the circular orbit limit by setting $e=0$ and come up with a simplified expression easily for $ \langle \dot{L}^{i}\rangle.$  But more importantly, from (\ref{dipoleang}) and (\ref{Ldot}) we can see, that the leading contribution to the angular-momentum flux comes from dipole term (at $\mathcal{O}(\frac{1}{c}$)) mentioned in (\ref{dipoleang}) \footnote{There is a $\mathcal{O}(\frac{1}{c^3})$ contribution coming from the scalar radiation also which can be traced back to the term (\ref{quadang}). }. This is consistent with the current results for scalar radiation. We get dominant contributions from the scalar radiation compared to the tensor part in both the energy and angular- momentum flux. Interested readers are referred to \cite{Zhang:2018prg}\footnote{Although one cannot compare our result directly with the result of \cite{Zhang:2018prg}, as they have done the analysis for a scalar-tensor theory which is only a subclass of Horndeski theory but a  qualitative similarity with our results exists.} .
\end{itemize}

\section{Orbit dynamics}\label{sec8}
As discussed in Sec.~(\ref{sec5}), we have tacitly been able to restore the inverse-square law in this non-GR theory. The advantage is that at the leading order, which is supposed to be the Newtonian order, the inverse law in this seemingly non-GR theory allows us to use the usual planetary laws of Kepler to study the orbit dynamics. There is a coupling between the scalar field and the gravitational field in this theory. So there must be a manifestation of such coupling in the dynamics of the binaries. At the Newtonian order, we saw such effects in Sec.~(\ref{sec5}) through a redefinition of the Gravitational constant, providing an effective gravitational constant as mentioned in (\ref{eq8}). Neglecting effects like radiation reaction on the system, we can write down the orbital energy and the orbital angular momentum of the system as
\begin{equation}
    E=-\frac{\mathcal{G}_{12}M\mu}{2a}\,,\quad L^{2}=\mathcal{G}_{12}M\mu^{2}a(1-e^{2})\,. \label{orbit}
\end{equation}
However, since the system is radiating energy we can expect a change of the orbit structures. We can however find this change by calculating the rate of change of energy and angular momentum of the system. Then we can write down change in semi-major axis and the eccentricity in the following way by taking time derivatives of (\ref{orbit}),
\begin{equation} \label{aedot}
    \langle \dot{a}\rangle=\frac{2a^{2}}{\mathcal{G}_{12}M\mu}\langle \dot {E}\rangle\,,\quad
    \langle \dot{e}\rangle=\frac{a(1-e^{2})}{\mathcal{G}_{12}M\mu\, e}\left[\langle \dot{E}\rangle-\frac{(\mathcal{G}_{12}M)^{1/2}}{a^{3/2}(1-e^{2})^{1/2}}\langle \dot{L}\rangle\right]\,.
\end{equation}
Then using (\ref{Edot}) and (\ref{Ldot}) we get from (\ref{aedot}), 
\begin{align}
    \begin{split}\label{aaedot} 
   \langle\dot{a}\rangle=    &  -\frac{2a^{2}}{\mathcal{G}_{12}M\mu}\Bigg[\frac{32\mu^{2} M^{3}\mathcal{G}_{12}^{3}}{5c^{5}G_{4(0,0)}a^{5}} \frac{1}{(1-e^{2})^{7/2}}\left(1+\frac{73}{24}e^{2}+\frac{37}{96}e^{4}\right)+\frac{2G_{4(1,0)}\xi}{G_{4(0,0)}}\Bigg( \frac{2A^{2}\mathcal{G}_{12}\mu^{2} M^{3}}{c\,a^{5}(1-e^{2})^{7/2}} e^{2}\left(1+\frac{e^{2}}{4}\right)\\& +\frac{\tilde{\alpha}^{2}}{3\,c^{3}a^{6}(1-e^{2})^{9/2}}\Big\{A^{2}_{d}a^{2}(1-e^{2})^{2} \left(1+\frac{e^{2}}{2}\right)+ \frac{\tilde{A}^{2}_{d}}{64}(64+608e^{2}+552e^{4}+36e^{6})+\\&  2A_{d}\tilde{A}_{d}\left(1+3e^{2}+\frac{3}{8}e^{4}\right)\Big\}+\frac{7\tilde{\alpha}^{2}A\tilde{A}\mu^{2}M}{6c^{3}a^{5}(1-e^{2})^{7/2}}e\Big(1+\frac{e^{2}}{7}\Big) +\frac{1}{60c^{5}}\frac{\tilde{\alpha}^{3}\mu^{2}\tilde{A}^{2}}{a^{5}(1-e^{2})^{7/2}}(160+424e^{2}+\frac{393}{8}e^{4})\\ & +\frac{2}{15c^{5}}\frac{\tilde{\alpha}^{3}\mu^{2}}{a^{5}(1-e^{2})^{7/2}}\Big\{A_{1}A_{d}\Big(1-\frac{132}{8}e^{2}-\frac{51}{8}e^{4}\Big)-A_{1}\tilde{A}_{d}\Big(1-\frac{2192}{64}e^{2}-\frac{3368}{64}e^{4}-\frac{279}{64}e^{6}\Big)\Big\} \Bigg) \Bigg]\,,
    \end{split}
\end{align} 
and 

\begin{align}
    \begin{split} \label{eedot}
        \langle\dot{e}\rangle= -&\frac{304}{15}\frac{\mu^{2} M^{3}\mathcal{G}_{12}^{2}}{\,c^5\,G_{4(0,0)}a^{5}} \frac{e}{(1-e^{2})^{5/2}}\left(1+\frac{121}{304}e^{2}\right)-\frac{a(1-e^{2})}{\mathcal{G}_{12}M\mu e}\Bigg[\frac{2\,G_{4(1,0)}\xi}{G_{4(0,0)}}\Bigg( \frac{2A^{2}\mathcal{G}_{12} \mu^{2} M^{3}}{c\, a^{5}(1-e^{2})^{7/2}} e^{2}\left(1+\frac{e^{2}}{4}\right)\\& +\frac{\tilde{\alpha}^{2}}{3c^{3}a^{6}(1-e^{2})^{9/2}}\Big\{A^{2}_{d}a^{2}(1-e^{2})^{2} \left(1+\frac{e^{2}}{2}\right)+ \frac{\tilde{A}^{2}_{d}}{64}(64+608e^{2}+552e^{4}+36e^{6})+\\&  2A_{d}\tilde{A}_{d}\left(1+3e^{2}+\frac{3}{8}e^{4}\right)\Big\}+\frac{7\tilde{\alpha}^{2}A\tilde{A}\mu^{2}M}{6c^{3}a^{5}(1-e^{2})^{7/2}}e\Big(1+\frac{e^{2}}{7}\Big) +\frac{1}{60c^{5}}\frac{\tilde{\alpha}^{3}\mu^{2}\tilde{A}^{2}}{a^{5}(1-e^{2})^{7/2}}(160+424e^{2}+\frac{393}{8}e^{4})\\ & +\frac{2}{15\,c^{5}}\frac{\tilde{\alpha}^{3}\mu^{2}}{a^{5}(1-e^{2})^{7/2}}\Big\{A_{1}A_{d}\Big(1-\frac{132}{8}e^{2}-\frac{51}{8}e^{4}\Big)-A_{1}\tilde{A}_{d}\Big(1-\frac{2192}{64}e^{2}-\frac{3368}{64}e^{4}-\frac{279}{64}e^{6}\Big)\Big\}\Bigg) \\ & -\frac{2\,(\mathcal{G}_{12}M)^{1/2}G_{4(1,0)}\xi}{3\,a^{3/2}(1-e^{2})^{1/2} G_{4(0,0)}}\Bigg( \frac{\tilde{\alpha}^{2}}{a^{6}c\,(1-e^{2})^{9/2}}\Big\{A^{2}_{d}(1-e^{2})^{2}+\tilde{A}^{2}_{d}(1+3e^{2}+\frac{3}{8}e^{4})+\\ & \frac{1}{2}A_{d}\tilde{A}_{d}(1-e^{2})(3+\frac{5}{4}e^{2})\Big\}+\frac{8}{75c^{3}}\frac{\mu^{2}\mathcal{G}_{12}^{3} m^{3}}{a^{7/2}(1-e^{2})}\left(1+\frac{7}{8}e^{2}\right) +\frac{\tilde{\alpha}^{5/2}A_{1}}{50c^{5}10\, a^{7/2}(1-e^{2})^{2}}\Big\{ A_{d}\left(1-\frac{7}{2}e^{2}\right)\\&-\frac{\tilde{A}_{d}}{{a^{4}(1-e^{2})^{4}}}\left(3-\frac{79}{2}e^{2}-\frac{81}{4}e^{4}\right)\Big\} \Bigg)  \Bigg]\,.
    \end{split}
\end{align}

The expressions above in combination with the Kepler-Law  $$a^{3}\left(\frac{2\pi}{P}\right)^{2}=\mathcal{G}_{12}M\,,$$ 
gives an off-hand estimate that the rate of change is somewhat related to the fraction of change in the orbital period of the binary \cite{Maggiore:2007ulw}. This provides an avenue to constrain some parameters of the theory as the orbital period of the binary is something we can measure accurately \cite{Maggiore:2007ulw}. Some observations can be made from the above expressions:

\begin{itemize}
    \item It can be easily seen from (\ref{aedot}) and (\ref{eedot}), the part coming from tensorial radiation for both   $\langle \dot{e}\rangle$ and $\langle \dot{a}\rangle$ is the of the same form as that of GR. We denote them as $\langle \dot{e}\rangle_T$ and $\langle \dot{a}\rangle_T.$
  \begin{align}
  \begin{split}
&\langle \dot{a}\rangle_T=  -\frac{64\,\mu M^{2}\mathcal{G}_{12}^{2}}{5c^{5}G_{4(0,0)}a^{3}} \frac{1}{(1-e^{2})^{7/2}}\left(1+\frac{73}{24}e^{2}+\frac{37}{96}e^{4}\right),\\&
\langle \dot{e}\rangle_T= -\frac{304}{15}\frac{\mu^{2} M^{3}\mathcal{G}_{12}^{2}}{\,c^5\,G_{4(0,0)}a^{5}} \frac{e}{(1-e^{2})^{5/2}}\left(1+\frac{121}{304}e^{2}\right).
\end{split}
  \end{align}
  
   \begin{figure}[t!]
  \centering
   \includegraphics[width=0.70\textwidth]{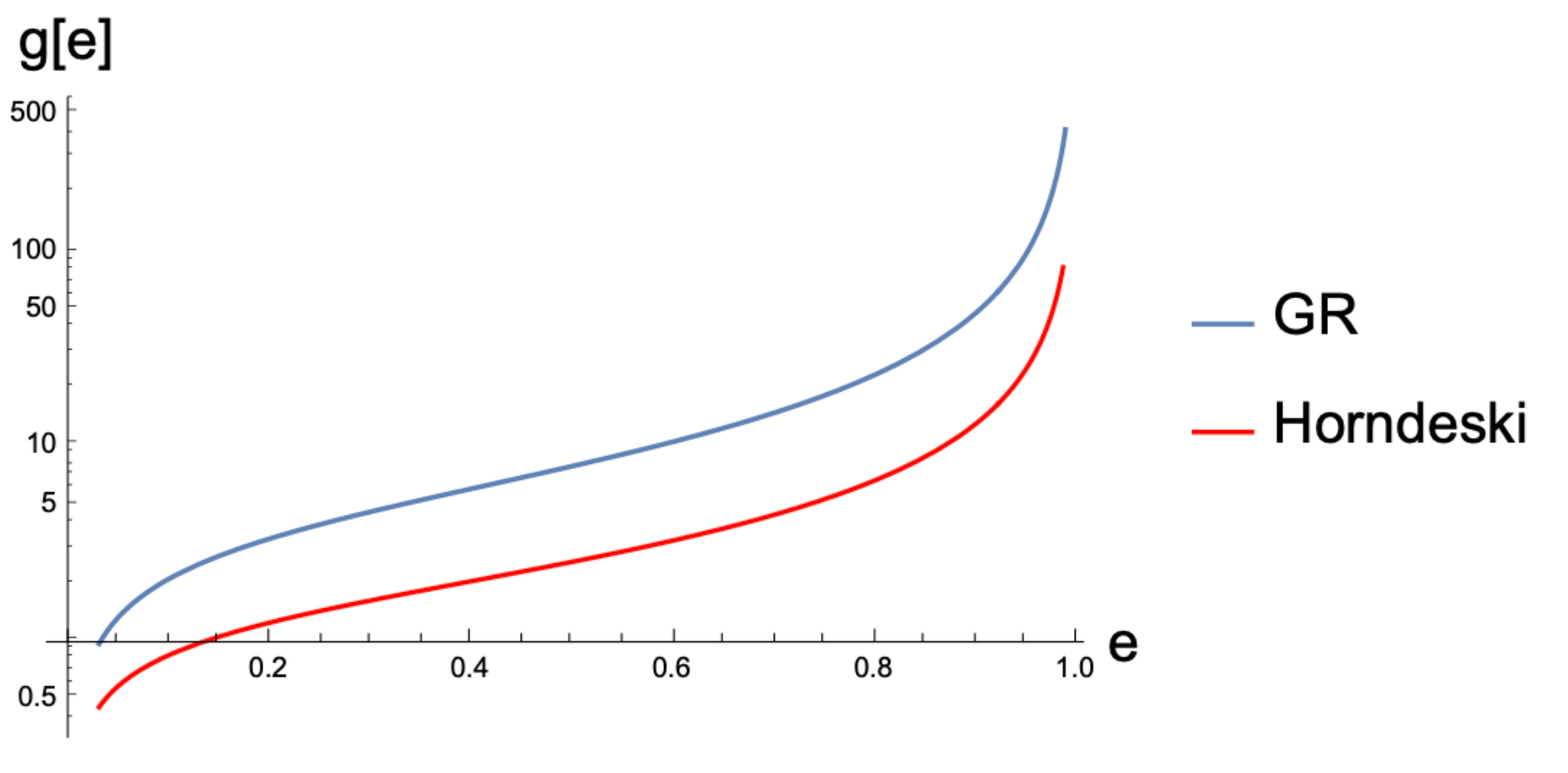} 
\caption{Figure shows the variation of $g(e)$ as defined in (\ref{ge}) with respect to the eccentricity, on a logarithmic scale. For convenience we have set various constant terms related the coupling constants, $c$ and mass of binaries  appearing in the expression of $a$ to be unity. One can use other choices for these parameters but this will not change the qualitative feature of the plot.}
 \label{fig:fig}
\end{figure}

  The effect of the Horndeski couplings is encoded in the effective gravitational constant $\mathcal{G}_{12}.$ This serves as consistency check of our computation. We can see that sign of both of these terms are negative which consistent with the existing results \cite{Maggiore:2007ulw,Alsing:2011er}. Hence the orbit will try to circularize. 
    \item Rest of the terms in (\ref{aaedot}) and (\ref{eedot}) comes from the scalar radiation part. Following our discussions in Sec.~(\ref{Energyflux}) and (\ref{angularmomentumflux}), we can easily see that the leading contribution comes from the scalar part compared to the tensor part. We know from (\ref{definition}) and (\ref{definition2}), that $A_d$  and $\tilde \alpha$ is independent of $c.$ Hence, the term appear at $\mathcal{O}(\frac{1}{c})$ in (\ref{eedot}) is the  leading order term  \footnote{ This $\mathcal{O}(\frac{1}{c})$ can be easily traced back to the $\mathcal{O}(\frac{1}{c})$ term appearing in (\ref{Ldot}) following our discussion below (\ref{Ldot}). } compared to the tensor part. This term comes from the second term of (\ref{Ldot}).
    \item To get a further insight into our results, we investigate the change in the orbital parameter, i.e. the semi-major axis, with respect to the eccentricity, as this gives physical insight into the orbital dynamics \cite{Maggiore:2007ulw}. Using (\ref{aedot}) and (\ref{eedot}) we apply chain rule to compute $\frac{da}{de}$ as a function of $e$ and then integrate it to get $a$ as function of $e.$ Following \cite{Maggiore:2007ulw}, we will get \begin{equation} a(e)=a_0\frac{g(e)}{g(e_0)},\label{ge}
    \end{equation}
    where $a_0$ is the initial value of the semi-major axis at some initial $e=e_0.$ In Fig.~(\ref{fig:fig}) we plotted this  function $g(e)$ \footnote{For GR, this function $g(e)$ is well known. Interested readers are referred to \cite{Maggiore:2007ulw} for more details. GR results can be recovered by the setting all the Horndeski couplings to zero in the expressions (\ref{aedot}) and (\ref{eedot}).}. In Fig.~(\ref{fig:fig}), the blue line indicates the result for GR, where we used the leading order post-Newtonian result for radiation coming entirely from the tensorial part. The red line represents Horndeski theory, for which the contribution comes from both the tensorial and scalar radiation parts. Note that, while making the plot, we have kept only the dipole contribution, which is the leading order contribution as discussed in Sec.~(\ref{Energyflux}) and Sec.~(\ref{angularmomentumflux}). One can use other terms in the scalar radiation as discussed in the Sec.~(\ref{Energyflux}) and Sec.~(\ref{angularmomentumflux}), but they will give subleading contributions, and the qualitative features of Fig.~(\ref{fig:fig}) will not change.\\

 \textit{Fig.~(\ref{fig:fig}) rightly shows that for a given $e$, the function $g(e)$ and hence the semi-major axis $a$ of the orbit due to radiation decreases more for the Horndeski theory compared to GR. This is in line with what we expect as the energy flux and angular momentum flux have extra contributions due to the scalar sector of the theory, which makes the binaries radiate away more energy than in GR.}
\end{itemize}


\section{Conclusion}\label{sec9}
In this paper, we have looked into the Horndeski theory without considering any screening. Our aim was to explore the orbital dynamics of such a theory and look into the radiation coming from eccentric binaries of such systems, making our analysis more general than the previous studies. The calculations have been done upto the next leading order Post-Newtonian expansion to find the solutions for the scalar field. The theory generically violates GWEP and admits a deviation from inverse-square law force expressions. However, we managed to restore the inverse law leading to usual planetary law relations. The radiations are obtained from an energy-momentum tensor obtained from short-wavelength approximation outlined in \cite{Isaacson:1968hbi}. We have also concluded that the orbital period change is related to the changes in the major axis and eccentricity of the orbit, which has a scalar and a tensor part. The tensor part is precisely like GR, and the only effect on the orbital period is the occurrence of an effective gravitational constant. Unlike the scalar sector, it has no monopole or dipole radiations. The monopole term in the scalar radiation vanishes for a quasi-circular case. However, since we have considered eccentric binaries, there's a non-zero monopole contribution in addition to the other terms. At this point appearance of monopole and dipole terms in the scalar radiation is mainly a byproduct of the computation. Perhaps it will be interesting to understand the origin of these terms from a more physical point of view along the lines of the discussion presented here \cite{Poisson:2003nc}. We leave this for future study.
Finally, we find that, for given value eccentricity $e,$ the semi-major axis $a$ of the orbit due to radiation decreases more for the Horndeski theory compared to GR. This is in line with what we expect as the energy flux and angular momentum fluxes have extra contributions due to the scalar sector of the theory, which makes the binaries radiate away more energy than that GR. This serves as a consistency check of our results.  
\par
It will be interesting to repeat this Post-Newtonian analysis by considering a more general matter Lagrangian \cite{Kobayashi:2019hrl}, by going beyond the point particle limit mentioned in (\ref{mact}). To carry out the computation explicitly, one might have to choose a specific form for the function $A(\phi)$ mentioned in (\ref{Horn}). We leave a detailed study of it for future investigation. Also, we have simplified our computation by setting the scalar field mass to zero. This has been done throughout the paper, enabling us to perform the orbital average for the eccentric orbit. It will be interesting in future to relax this assumption and figure out how to perform the orbital average. \par
But even after these simplifying assumptions, the expressions mentioned in (\ref{Edot}), (\ref{Ldot}), (\ref{aedot}) and (\ref{eedot}) can be utilized to explore the theory-finding constraints on the parameters of the theory just as done in  using eccentric binary pulsar data along the lines of \cite{Gupta:2021vdj,Anderson:2019eay,Guo:2021leu}.
Furthermore, our analysis can be utilized to look at the imprint of the signatures of such theories in GW chirping. Recent GW observations offer an excellent possibility to test such theories. The interpretation of such observations relies on the orbital dynamics in such theories, which is where our analysis becomes significant. One can integrate these flux expressions mentioned in (\ref{Edot}), (\ref{Ldot}), (\ref{aedot}) and (\ref{eedot})and find the expression of the phase that enters into the exponent of the gravitational waveform \cite{Ma:2019rei}. This phase then contains information about the Horndeski coupling and the eccentricity parameter. Then one can do a proper statistical analysis to constrain the coupling parameters and search for the signature of the presence of eccentricity. Note that, as discussed in the Sec.(\ref{Energyflux}) and Sec.(\ref{angularmomentumflux}) we get a leading order (compared to the tensorial part) contribution due to the presence of the dipolar radiation. So our leading order PN results should be enough for future investigation as that will already have contributions from some of the Horndeski couplings. We hope to report it in an upcoming publication \cite{Upc}.\par
Also, it may be the case that the effect of the Horndeski theory at 1-PN order in the presence of eccentricity may be mimicked by other theories or even GR without eccentricity but with higher-order PN terms and vice versa in the phase of GW waveform. In that context, perhaps one can try to explore the \textit{odds ratios} along the lines of \cite{Thrane:2018qnx} to see which theoretical waveform coming from different competing theories fits better with a particular GW signal.  Also, one can look into search algorithms by constructing appropriate template banks of this theory and provide a statistical likelihood for finding such binaries. We would also like to look into the theory in the presence of screening mechanisms which will be the most general scenario. Furthermore, one might also consider the complete set of post-Keplerian parameters and explore the sensitivities of such theories to see whether they can be related to some astrophysical objects like its dependence on the star mass in this gravity theory. These will provide an exciting avenue to explore. 

\section*{Acknowledgements}

Research of A.C. is supported by the Prime Minister's Research Fellowship (PMRF-192002-1174) of Government of India. A.B is supported by Start-Up Research Grant (SRG/2020/001380), Mathematical Research Impact Centric Support Grant (MTR/2021/000490) by the Department of Science and Technology Science and Engineering Research Board (India) and Relevant Research Project grant (202011BRE03RP06633-BRNS) by the Board Of Research In Nuclear Sciences (BRNS), Department of atomic Energy, India. 

\begin{appendix}
\section{Multipole moments} \label{A}

We list below the expressions for time derivative of the individual multipole moments used in the calculation of fluxes in the main text. 
\begin{align}
    \begin{split}
  &      \dot{\mathcal{M}}_{0}=-A\,\mu\, M \frac{e\sin\theta}{a(1-e^{2})}\sqrt{\frac{\mathcal{G}_{12}} m}{a^3}(1-e^{2})^{-3/2}(1+e\cos\theta),
    \end{split}
\end{align}
\begin{align}
    \begin{split}
  &  \ddot{\mathcal{M}}^{k}_{1}=-\frac{\tilde{\alpha}\,\mu\, (\alpha_{1}-\alpha_{2})}{(1-e^{2})^{3}}a(1-e^{2})\\ & \left\{\cos\left[\omega\left(t-\frac{r}{c}\right)\right]\left(1+e\cos\left[\omega\left(t-\frac{r}{c}\right)\right]\right)^{2},\, 
  \sin\left[\omega\left(t-\frac{r}{c}\right)\right]\left(1+e\cos\left[\omega\left(t-\frac{r}{c}\right)\right]\right)^{2},0\right\}\\ & +\frac{\mu}{c^{2}}\left[-\left(\frac{3}{G_{4(0,0)}}-6\frac{G_{4(1,0)}}{G_{4(0,0)}}\xi+\gamma_{1}\right)m_{2}\beta_{2}+\left(\frac{3}{G_{4(0,0)}}-6\frac{G_{4(1,0)}}{G_{4(0,0)}}\xi+\gamma_{2}\right)m_{1}\beta_{1}\right]\\ & \Big\{\cos\left[\omega\left(t-\frac{r}{c}\right)\right]+e\cos^{2}\left[\omega\left(t-\frac{r}{c}\right)\right]-2e\sin^{2}\left[\omega\left(t-\frac{r}{c}\right)\right] \left(1+e\cos\left[\omega\left(t-\frac{r}{c}\right)\right]\right)^{3}, \\ & -\sin\left[\omega\left(t-\frac{r}{c}\right)\right]\Big(1+3e\cos\left[\omega\left(t-\frac{r}{c}\right)\right]\Big)\left(1+e\cos\left[\omega\left(t-\frac{r}{c}\right)\right]\right)^{3},0\Big\},
    \end{split}
\end{align}
\begin{align}
    \begin{split}
      &  \dddot{\mathcal{M}}^{11}_{2}=\tilde{\alpha}^{3/2}\frac{\mu(1-\tilde{\alpha})a^{2}}{(1-e^{2})^{5/2}}(4+3e\cos\theta)\sin2\theta(1+e\cos\theta)^{2},\\&
       \dddot{\mathcal{M}}^{22}_{2}=-\tilde{\alpha}^{3/2}\frac{\mu(1-\tilde{\alpha})a^{2}}{(1-e^{2})^{5/2}}(8\cos\theta+e(5+3\cos2\theta))\sin2\theta(1+e\cos\theta)^{2},\\&
        \dddot{\mathcal{M}}^{11}_{2}=-\tilde{\alpha}^{3/2}\frac{\mu(1-\tilde{\alpha})a^{2}}{(1-e^{2})^{5/2}}(5e\cos\theta+8\cos2\theta+3e\cos3\theta)(1+e\cos\theta)^{2}, \nonumber
    \end{split}
\end{align}
\begin{align}
    \begin{split}
     \ddddot{\mathcal{M}}^{1ll}_{3}=  & -\frac{1}{4}\tilde{\alpha}^{2}\frac{\mu^{3}A_{1}a^{3}}{(1-e^{2})^{3}}((-4+29e^{2})\cos\theta +5e(8\cos2\theta+3e\cos3\theta))(1+e\cos\theta)^{2},
    \end{split}\nonumber
\end{align}
\begin{align}
    \begin{split}
  \ddddot{\mathcal{M}}^{1ll}_{3}=   -\frac{1}{2}\tilde{\alpha}^{2}\frac{\mu^{3}A_{1}a^{3}}{(1-e^{2})^{3}}(-2+27e^{2} +5e(18\cos\theta+3e\cos2\theta))(1+e\cos\theta)^{2}.
    \end{split}
\end{align}
Here $\alpha_1,\alpha_2,\tilde \alpha, A_1$ and $A$ are defined in (\ref{definition1}), (\ref{definition}) and (\ref{definition2}). The effective gravitational constant $\mathcal{G}_{12}$ is defined in (\ref{effectivegg}).

\end{appendix}

\end{document}